\documentclass{hep99}
\begin{document}

\title{Hadron yields from thermalized minijets at RHIC and LHC}

\author{N. Hammon$^1$, H. St\"ocker$^1$, W. Greiner$^1$ and A. Dumitru$^2$}
%

\address{$^1$ Institut F\"ur Theoretische Physik, Robert-Mayer Str. 10, 60054 Frankfurt am Main, Germany\\
$^2$ Physics Dept., Columbia University, 704 Pupin Hall, Mail Code 9318, 538W 120th Street, New York, NY 10027, USA}

\abstract{We calculate the yields of pions, kaons, and $\phi$-mesons for RHIC and LHC
energies assuming thermodynamical equilibration of the produced 
minijets, and using as input results from pQCD for the energy densities
at midrapidity.
In the calculation of the production of partons and of
transverse energy one has to account for nuclear shadowing. By using
two parametrizations for the gluon shadowing one derives energy densities
differing strongly in magnitude.
In this publication we link those perturbatively calculated energy densities of
partons via entropy conservation in an ideal fluid to the hadron multiplicities
at chemical freeze-out.}

\maketitle

\fntext{1}{E-mail: hammon@th.physik.uni-frankfurt.de, stoecker@th.physik.uni-frankfurt.de, greiner@th.physik.uni-frankfurt.de}
\fntext{2}{E-mail: dumitru@mail-cunuke.phys.columbia.edu}

\section{Introduction}
Particle production in high-energy heavy-ion reactions at the BNL-RHIC
and CERN-LHC colliders will soon provide interesting insight into nuclear
modifications of semi-hard processes
\cite{Mueller}. This is because
pQCD processes involving gluons in the initial state may dominate the
inelastic AA cross-section at collider energies.
In particular, they might lead
to a better understanding of the gluon
distribution in large nuclei, which is not accessible in DIS.

In \cite{acapella} the effect of nuclear
shadowing of the parton distribution functions on the charged particle
multiplicity at midrapidity
has been investigated assuming no rescattering between the produced minijets
and the hadrons they fragment into. Here, we will take the opposite point of
view and assume maximal rescattering, i.e.\ local thermal and chemical
equilibrium of the minijets. We compute final-state hadron multiplicities
of various hadron species under the assumption of entropy conservation.

In \cite{ham99} we calculated the initial conditions at RHIC and LHC
by means of pQCD above the semihard scale $p_T=2$ GeV to derive the 
number and energy densities of partons at midrapidity. In that calculation we
explicitely
included the shadowing effect on the parton distribution functions entering the
formulas for the production of flavor $f=g, q, \bar q$ in the minijet approach.
We employed two different parametrizations for the shadowing effect accounting 
for weak and strong gluon shadowing, respectively, \cite{ham99, eskola3}.
A direct consequence of the
shadowing effect is the decrease in the production of partons of given
momentum $p_T$, i.e.~a decrease of transverse energy production at midrapidity.
We calculate the first $E_T$ moment with and without shadowed pdf's   
and with a cut-off function $\epsilon (y)$ ensuring that we only count
scatterings
into the central rapidity region ($\left| y\right| \leq 0.5$):
\begin{eqnarray}
&& \sigma^{f}_{hard} \left < E_T\right >_{hard} = \int dE_T \frac{d\sigma ^f}{dE_T}
\left < E_T\right > \\
&& = \int dp_{T}^{2} dy_1 dy_2 \sum_{ij,kl} x_1 f_{i}(x_1,Q^2)~ x_2 f_{j} (x_2,Q^2)\nonumber \\
&&\left [ \delta_{fk} \frac{d\hat \sigma^{ij\rightarrow kl}}{d\hat t}
(\hat t, \hat u)+
\delta_{fl} \frac{d\hat \sigma^{ij\rightarrow kl}}{d\hat t}(\hat u, \hat t)
\right ]
\frac{p_T\epsilon (y)}{1+\delta_{kl}}\nonumber.
\end{eqnarray}
Motivated by the factorization in QCD one can assume that
the production of transverse energy in AA collisions can be split up into a
hard and a soft contribution as
\begin{equation}
\bar{E}_T (b) = T_{AA}(b)\left[ \sigma_{hard}^{pp}\left<E_T\right>_{hard}^{pp}+
\sigma_{s}^{pp}\left<E_T\right>_{s}^{pp} \right].
\end{equation}
With an energy independent value of $\sigma_{s}^{pp}=32$ mb one derives
\cite{eskola2} $\sigma_{s}^{pp}\left<E_T\right>_{s}^{pp}=15$ mb GeV.   
With $T_{AuAu}=29/$mb one can derive the soft
contribution (i.e.~the one for $p_T\leq 2$ GeV) to the energy density for RHIC
as $\varepsilon _{soft} = 33.7~{\rm GeV/fm}^{3}$.
When comparing to the soft contributions one finds with
$\sigma_{hard}\left<E_T\right>_h = \left(\sigma^g + \sigma^q +
\sigma^{\bar q}\right)_{hard}\left<E_T\right>_{hard}$
that the ratio of soft to hard contribution
\begin{equation}
R_{sh}=\frac{\sigma_{soft}\left<E_T\right>_s}{\sigma_{hard}\left<E_T\right>_h}
\end{equation}
is $R_{sh}=0.47$ for no, $R_{sh}=0.77$ for strong, and $R_{sh}=0.47$ for weak shadowing,
respectively, since $\varepsilon = 71.5$~{\rm GeV/fm}$^{3}$ for no, 
$\varepsilon = 43.6$~{\rm GeV/fm}$^{3}$ for strong, and
$\varepsilon = 71.9$~{\rm GeV/fm}$^{3}$ for weak gluon shadowing .
This implies that at RHIC the soft component could
even dominate if it were energy independent and also unaffected by the
shadowing effect.\\

For LHC we only calculated the contribution of the gluons that
strongly dominate all partonic processes due to the large distribution function at
small momentum fractions.
The results for the energy densities for no, strong, and weak gluon
shadowing, then are
$\varepsilon _g = 1229.7~{\rm GeV/fm}^{3}$, $\varepsilon _g = 144.8~{\rm GeV/fm}^{3}$,
and $\varepsilon _g = 678.6~{\rm GeV/fm}^{3}$.
With $T_{PbPb}=32/$mb one can derive the energy density from the soft part
and has $\varepsilon _{soft} = 35.8~{\rm GeV/fm}^{3}$
which is slightly larger than at RHIC due to the larger nuclear overlap function,
i.e.~the larger number of effective scatterings in the Glauber picture leading to
the transverse energy production.
The relative weight
\begin{equation}   
R_{sh}=\frac{\sigma_{soft}\left<E_T\right>_s}{\sigma_{hard}\left<E_T\right>_h}
\end{equation}
between soft and hard contributions therefore changes and becomes
$R_{sh}=0.029$ for no, $R_{sh}=0.25$ for strong, and $R_{sh}=0.052$ for weak shadowing,
respectively.\\
Therefore the soft contributions gain much more weight in this naive picture
due to the strong effect of shadowing on the small-$x$ gluons
(for further details of the calculation,
the shadowing parametrizations, etc.~see \cite{ham99}).

\section{Hadron Multiplicities at Chemical Freeze-Out}
Having calculated the energy densities at midrapidity for RHIC and LHC in pQCD
we connect $\varepsilon _i$ to the number of hadrons at midrapidity by assuming
an ideal fluid that is characterized by entropy conservation from the  
quark-gluon plasma to
the hadron gas, $dS_i/dy = dS_f/dy$ \cite{Bjorken}.
We relate the energy density of the quark-gluon plasma to its
entropy density via the bag model equation of state \cite{cho}.
We account for $u$, $d$, $s$ quarks (with masses $m_u=m_d=0$, $m_s=150$ MeV),
the antiquarks, and gluons.
The total produced entropy $dS_i/dy$ is obtained from the entropy density   
at midrapidity as $dS_i/dy = V_i s_i$ with the initial volume of the central
region $V_i =\pi R_{A}^{2}\tau_i$,
which is numerically $V_i=12.9~{\rm fm}^3$ for $Au+Au$ and
$V_i=13.4~{\rm fm}^3$ for
$Pb+Pb$ at $b=0$ with $\tau =0.1~{\rm fm}/c$. Since we assumed an ideal
fluid the
total entropy is
conserved throughout the expansion until freeze-out which is chosen here to   
happen at a temperature $T_{FO}=160$ MeV. For simplicity we furthermore assume
vanishing chemical potentials in the central rapidity region, i.e.~that all
conserved currents are identically zero. If this were not true one would
have to multiply by factors ${\rm exp}(\mu_i/T)$ (in Boltzmann approximation).
The entropy density of the hadronic fluid is calculated assuming an ideal gas
composed of all hadrons up to a rest-mass of 2 GeV. Their respective
occupation numbers are given by Fermi-Dirac or Bose-Einstein distribution
functions, respectively. Thus, $T_{FO}$ and $dS_f/dy=dS_i/dy$ determine the
multiplicity of each hadron species uniquely \cite{pbm, adrian}.
Feeding from post freeze-out decays of heavier resonances is also taken
into account.
\section{Results}
With the model outlined above and the energy densities derived
above we calculated the number of a variety of hadrons
at midrapidity. We also include the multiplicities due to the soft contributions
and quote the initial temperatures for a QGP of three flavors.
For LHC we derived the yields shown in table \ref{table4} and for RHIC the ones in
table \ref{table5}.
\begin{table}
\begin{tabular}{|c||c|c|c|c|} \hline
{\bf LHC} & $\pi^\pm$ & K & $\phi$ & $T_i$ \\
\hline \hline
no shad. & 2680 & 478 & 32.1 & 881 MeV\\ \hline
weak shad. & 1720 & 306 & 20.6 & 760 MeV \\ \hline
strong shad. & 538 & 95.9 & 6.5 & 516 MeV\\ \hline
soft contrib. & 187 & 33 & 2 &   \\ \hline
\end{tabular}
\caption{\it Hadron yields at freeze-out with initial conditions from pQCD for LHC.
The initial temperatures are calculated from the energy densities produced via hard
processes and are for a three flavor quark-gluon plasma with
\protect$m_u=m_d=0$ {\rm MeV} and \protect$m_s=150$ {\rm MeV}, respectively.}
\label{table4}
\end{table}   
For the latter one, one can clearly see
that there is no change in the hadron yield for weak shadowing and the unshadowed case,
respectively, due to the almost identical energy density serving as an input for the   
calculation.
\begin{table}
\begin{tabular}{|c||c|c|c|c|} \hline
{\bf RHIC} & $\pi^\pm$ & K & $\phi$ & $T_i$ \\
\hline \hline
no shad. & 316 & 56.3 & 3.8 &  433 MeV\\ \hline
weak shad. & 316 & 56.3 & 3.8 &  433 MeV \\ \hline
strong shad. & 217 & 38.7 & 2.6 &  383 MeV\\ \hline
soft contrib. & 179 & 32 & 2 & \\ \hline
\end{tabular}
\caption{\it Hadron yields at freeze-out with initial conditions from pQCD for RHIC.}
\label{table5}
\end{table}   
\section{Conclusions and outlook}
We computed the rapidity densities
of a variety of hadrons
based on the assumption of entropy conservation of an ideal fluid. The
entropy densities were derived from the energy densities which in turn were
caluclated by means of pQCD \cite{ham99}. We find that in the limit that the minijets
equilibrate locally the effect of shadowing on the
hadron yield is not as large as on the pure partonic degrees of freedom.
This can be seen e.g.~in the ratio of energy densities between unshadowed
and strongly shadowed gluons at LHC, which is about a factor of 9, while the ratio of
the hadron yields is only a factor of 5. Since vanishing net baryon and
strangeness densities were assumed,
the relative depletion of shadowed to unshadowed gluon distribution
is independent of the particle species. For RHIC, even without shadowing
we only get about 300 pions since the perturbative calculation, entering
via the energy densities, was performed with the cut-off $p_T =2$ GeV. Therefore,
the soft contribution constitutes a significant part of the transverse energy
and therefore of the particle multiplicities \cite{adrian}.
\centerline{\bf Acknowledgements}
This work was supported by BMBF, DFG, and GSI. 

\end{document}